# Step and Search Control Method to Track the Maximum Power in Wind Energy Conversion Systems – A Study

V.Karthikeyan[1], V.J.Vijayalakshmi[2], P.Jeyakumar[3]

**Abstract –** *A simple step and search control strategy for extracting maximum output power from grid connected Variable Speed Wind Energy Conversion System (VSWECS) is implemented in this work. This system consists of a variable speed wind turbine coupled to a Permanent Magnet Synchronous Generator (PMSG) through a gear box, a DC-DC boost converter and a hysteresis current controlled Voltage Source Converter (VSC). The Maximum Power Point Tracking (MPPT) extracts maximum power from the wind turbine from cut-into rated wind velocity by sensing only by DC link power. This system can be connected to a micro-grid. Also it can be used for supplying an isolated local load by means of converting the output of Permanent Magnet Synchronous Generator (PMSG) to DC and then convert to AC by means of hysteresis current controlled Voltage Source Converter (VSI).*

## I. Introduction

Wind energy is proving itself as a cost of effective and reliable energy resource around the world. Grid connected, constant speed wind energy systems employing induction generator are popular and they extract optimum power from the wind for a single wind speed. The Variable Speed Wind Energy Conversion System (VSWECS) integrated with power electronics due to their capability of extracting energy capture, reduced mechanical stresses and aerodynamic noises. For optimum energy extraction, the speed of the turbine should be varied with wind speed so that the optimum tip-speed ratio is maintained [1]. In this system, the generator directly couples the grid to drive turbine. In Variable Speed Wind Energy Conversion System, the rotor is permitted to rotate at any speed by introducing power electronic converters between the generators and the grid. The construction and performance of this system are very much dependent on the mechanical subsystems, pitch control constant, etc., with variable speed, the result is mechanical stresses that decrease the power quality. Furthermore, with constant speed there is only one wind velocity that results in an optimum tip-speed ratio [2]. Wind energy is ecological forthcoming, unlimited, secure and competent of supply substantial amount of power. However, outstanding to wind's unpredictable environment, clever manage strategies must be put into practice to yield as a large amount probable wind energy as likely as it is obtainable. Research to extract maximum power out of wind energy is an essential part of making wind energy much more viable and attractive [3]. Best use of the wind resources imposes the requirement of being constantly operating the wind turbine generator near the Maximum Power Point (MPP) independently of the wind speed conditions. Thus, the relationships between power generation costs and investment costs of Wind Energy Conversion System (WECS) are minimized. This work proposes a fuel detailed modeling and a novel control scheme of a three-phase grid-connected WECS. The WECS model consists of a variable speed wind turbine generator and the electronic power conditioning system is composed of a back-to-back AC-to-DC-to-AC power converter. The control consists of a multi-level hierarchical structure and incorporates a Maximum Power Point Tracking (MPPT) for best use of a permanent magnet generator is connected to a fixed DC-link through a 3-phase rectifier and a step-up DC-DC converter, which permits the implementation of the MPPT. A Pulse Width Modulation (PWM) voltage source converter is used to convert the energy produced by wind turbines into useful electricity and to provide requirements for power grid interconnection [4]. In this paper, a simple step and search control strategy to extract the maximum power from the grid connected VSWECS is implemented. Unlike other control strategies reported in the literature, this is achieved without the knowledge of turbine parameters.

## II. Modeling of Variable Speed Wind Energy Conversion System (VSWECS)

The wind generator is formed by a fixed pitch wind turbine, a permanent magnet synchronous generator, a passive rectifier, a DC-DC boost converter and a hysteresis current controlled Voltage Source Converter (VSC). Wind energy can be harnessed by a wind energy conversion system, composed of wind turbine blades, an electric generator, a power electronic converter and the corresponding control system. There are different WECS configurations based on using synchronous or asynchronous machines and stall-regulated or pitch regulated systems. However, the functional objective of these systems is the same: converting the wind kinetic energy into electric power and injecting this electric power into a utility grid. For charging the kind of maker in WECS, criteria such as prepared individuality, heaviness of lively resources, value, preservation aspects and the suitable kind of control electronic converter are used.

### II.1. Wind Turbine Characteristics

Wind turbine is characterized by the non-conventional curve of co-efficient of performance $C_p$ as a function of tip-speed ratio λ. It can be defined as follows [5]

$$\lambda = \frac{r\omega}{v} \quad (1)$$

The power co-efficient $C_p$ varies with λ. For the wind turbine used for this study, $C_p$ as a function of λ is expressed by the following equation (2) and is shown in Fig.2.

$C_p = 0.043 - 0.108\lambda + 0.146\lambda^2 - 0.0602\lambda^3 + 0.0104\lambda^4 - 0.0006\lambda^5$ (2)

The output power of the wind turbine $P_t$ may be calculated as,

$$P_t = \frac{1}{2} C_p(\lambda) \ell A v^3 \quad (3)$$

It can be observed that $C_p$ maximum when λ is equal to 7.5. This value is higher than Belz limit, because of polynomial approximation of the actual characteristics. The torque developed by the wind turbine can be expressed as,

$$T_t = \frac{P_t}{\omega} \quad (4)$$

Combining equations (4), (3) & (1), the expression for the torque may be written as,

$$T_t = \frac{1}{2} \ell A r \frac{C_p(\lambda)}{\lambda} v^2 \quad (5)$$

The power extracted from the wind is maximized when the power co-efficient $C_p$ is at its maximum. This occurs at a defined value of the tip-speed ratio λ. Hence for each wind speed there is an optimum rotor speed where maximum power is extracted from the wind. Therefore if the wind speed is assumed to be constant, the value of $C_p$ depends on the wind turbine rotor speed. Thus by controlling the rotor speed, the power output of the turbine is controlled.

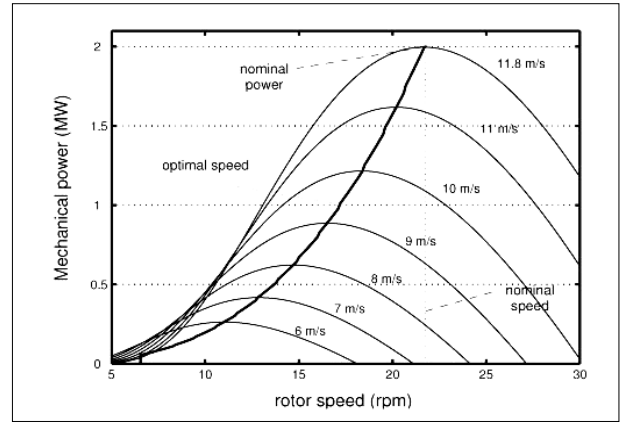

Fig.1.Typical power versus speed characteristics of a wind turbine

## III. Permanent Magnet Synchronous Generator (PMSG)

Permanent magnet generators are widely used for the portable electric source driven by engines. It is required that the construction of generator is simple and robust from the view points of easy maintenance and high reliability [6]. From all the generators that are used in wind turbines the PMSG's have the highest advantages because they are stable and secure during normal operation and they do not need an additional DC supply for the excitation circuit.

Initially used only for small and medium powers the PMSG's are now used also for higher powers [7]. The equations for modeling a PMSG, using the motor machine conversions are given by,

$$\frac{di_d}{dt} = \frac{1}{L_d}[u_d + Pw_g L_q i_q - R_d i_d] \quad (6)$$

$$\frac{di_q}{dt} = \frac{1}{L_q}[u_q - Pw_g(L_d i_d + M i_f) - R_q i_q]$$

Where $i_d$, $i_q$ are the stator currents, $u_d$, $u_q$ are the stator voltages, P is the no. of pairs of poles, $L_d$, $L_q$ are the stator inductances, $R_d$, $R_q$ are the stator resistances, M is the mutual inductance, $i_f$ is the equivalent rotor current. In order to avoid demagnetization of permanent magnet in the PMSG, a null stator current associated with the direct axis is imposed [8].

*III.1. Pulse Width Modulation (PWM) Rectifier*

Each switch of PWM converter is represented as binary resistance. The value of this resistance is infinite if the switch is "OFF" and zero if it is "ON". The switching functions for the rectifier are derived by a controller which maintains dc link voltage constant while shaping the phase current as the phase voltage [5]. It employs a very simple control structure, where a proportional-integral (PI) controller is used to regulate the dc output voltage and provide the magnitude value of the sinusoidal reference currents. The inputs of this controller is the error between the square of the reference dc voltage and the dc bus capacitors voltage to have a linear adjustment of the output dc average power, the output is considered as the magnitude of the desired supply currents and the reference currents are estimated by multiplying this magnitude with the unit sine vectors in phase with the ac main source voltages provided by the phase-locked-loop (PLL) [9].

*III.2. Pulse Width Modulation (PWM) Inverter*

The Procedure for modeling PWM inverter is same as that of the PWM rectifier. The reference currents required for the PWM inverter are generated by a separate controller. This controller determines the magnitude of the reference current in such a way that the inverter output power is equal to the reference power that is set by MPPT. These reference currents are in phase with the respective grid voltages. This is achieved by a 3-phase PLL.

In this paper, a VSWECS consisting of PMSG, two PWM converters and a MPPT is considered. Fig.2. shows the schematic diagram of the system. Output power from the permanent magnet generator is first converted into dc. The output voltage is rectified using a three-phase diode bridge rectifier. The dc-to-dc converter is used to control the dc voltage $V_{dc}$ across capacitor $C_1$. Both the conversions are performed at unity power factor and the dc link voltage is kept constant. Maximum power point tracking extracts optimum power from the wind turbine for the wind speeds cut-in to rated, by generating suitable reference current to the inverter. The result is fed into a PI controller whose output is compared to a triangular waveform to determine when to turn the DC-to-DC boost converter switch ON or OFF. The hysteresis current controlled VSC interfaces the wind turbine system with the power grid.

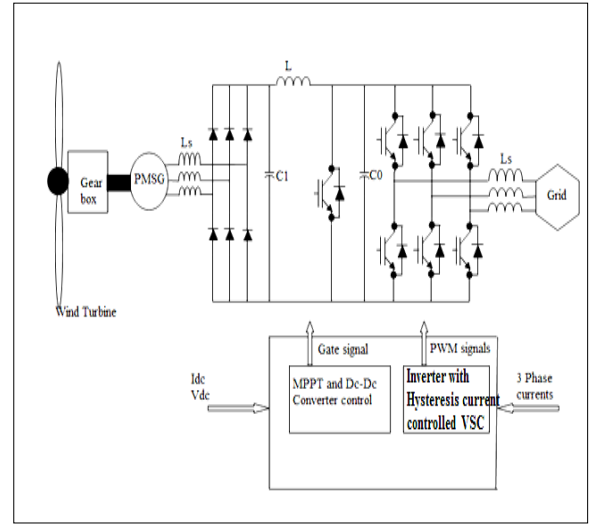

Fig.2. WECS with MPPT and Hysteresis current controlled VSC

### IV. Control Strategy

*IV.1. Maximum Power Point Tracking Algorithm*

The MPPT process in the proposed system is based on directly adjusting the dc-to-dc converter duty cycle according to the result of the comparison of successive wind turbine output power measurements. Although the wind speed varies highly with time, the power absorbed by the wind turbine varies relatively slowly, because of the slow dynamic response of the interconnected wind turbine system. Thus, the problem of maximizing the wind turbine output power using the converter duty cycle as a control variable can be effectively solved using the steepest ascent method. In equation (3), substituting value of v from (1) we have,

$$P_t = \frac{1}{2} \ell C_p \pi \frac{r^5 \omega^3}{\lambda^3} \quad (8)$$

At optimum value of $C_p$, $\lambda = \lambda_{opt}$, hence optimum output power of the wind turbine is,

$$P_{opt} = \frac{1}{2} \ell C_{p(opt)} \pi \frac{r^5 \omega^3}{\lambda_{opt}^3} = k\omega^3 \quad (9)$$

Hence by measuring the rotor speed of the wind turbine, it is possible to set the reference power, which will make $C_p = C_{p\ (opt)}$.

Considering the wind turbine power characteristics, it is obvious that at the points of maximum power production is [10],

$$\frac{dp}{d\Omega} = 0 \quad (10)$$

Where $\Omega$ is the wind turbine rotor speed

Applying the chain rule, the above equation can be written as,

$$\frac{dp}{d\Omega} = \frac{dp}{dD} \cdot \frac{dD}{dV_{WG}} \cdot \frac{dV_{WG}}{d\Omega_e} \cdot \frac{d\Omega_e}{d\Omega} = 0 \quad (11)$$

Where, $V_{WG}$ is the rectifier output voltage level; $\Omega_e$ is the generator phase voltage angular speed.

In case of boost-type dc-to-dc converter, its input voltage is related to the output voltage and duty cycle as follows

$$D = \frac{V_o}{V_{WG}} \quad (12)$$

Where $V_0$ is the voltage level

$$\frac{dD}{dV_{WG}} = \frac{-1}{V_{WG}^2} V_0 \neq 0$$

Then,

$$\frac{dP}{d\Omega} = 0 \Leftrightarrow \frac{dP}{dD} = 0 \quad (13)$$

Thus, the function P (D) has a single extreme point, coinciding with the wind turbine maximum power point, and the dc-to-dc converter duty cycle adjustment according to the control law of ensures convergence to the wind turbine maximum power point under any wind speed condition. The power maximization process is shown in fig.3. Since the power-speed plot is shown for three different wind speeds, where $v_1 < v_2 < v_3$. The arrows show the trajectory in which the turbine will be operated using the maximum power tracking algorithm explained above. If the wind speed is $v_1$, the controller will search for the maximum power. If the wind changes to $v_3$ the turbine is no longer being operated at the maximum power point so the controller will search for the new maximum power point. After reaching the maximum point it will operate the wind turbine at the optimal point until wind changes, thus searching for maximum power at any wind speed. In order to optimize the maximum power search algorithm presented above, a step that combines speed of convergence and accuracy of results was developed. The variable step method is based on the Newton-Raphson method. In order to search for maximum power at any wind speed four conditions must be met. If P (k) ≥ P (k −1) and $V_{dc}$ (k) −$V_{dc}$ (k−1), the dc side voltage reference need to be increased by $\Delta V_{dc}$. This condition is met when the turbine operates on the low speed side of the power curve, shown on Fig. 3. If P (k) ≥ P (k −1) and $V_{dc}$ (k) < $V_{dc}$ (k −1), the wind turbine is being operated in the high speed side and the dc reference voltage needs to be decreased by $\Delta V_{dc}$. When P (k) < P (k −1) and $V_{dc}$ (k) ≥ $V_{dc}$ (k −1), the maximum power point is passed and a step back must be taken, decreasing the reference voltage by $\Delta V_{dc}$. When P (k) < P (k −1) and $V_{dc}$ (k) < $V_{dc}$ (k−1), the power is decreasing on the low speed side, therefore the voltage reference is to be increased by $\Delta V_{dc}$.

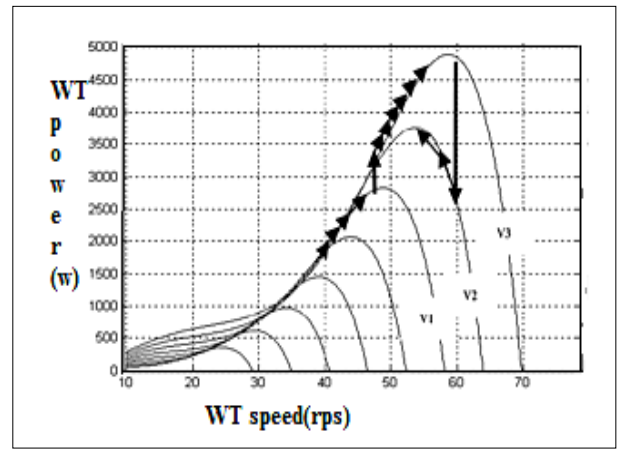

Fig.3. MPPT Process

### IV.2. Hysteresis Current controlled Voltage Source Converter (VSC).

Among the various PWM techniques, the hysteresis current control is used very often because of its simplicity of implementation [11]. Also besides fast response current loop, the method does not need any knowledge of load parameters. A hysteresis control method provides good large-signal response and stability. The basic implementation of hysteresis current control is based on deriving the switching signals from the comparison of the current error with a fixed tolerance band. This control is based on the comparison of the actual phase current with the tolerance band around the reference current associated with that phase. This is mainly due to the interference between the commutations of the three phases, since each phase current not only depends on the corresponding phase voltage but is also affected by the voltage of the other two phases. Depending on load conditions switching frequency may vary during the fundamental period, resulting in irregular inverter operation. The current error is defined as,

$$i_e = i - i_{ref} \qquad (14)$$

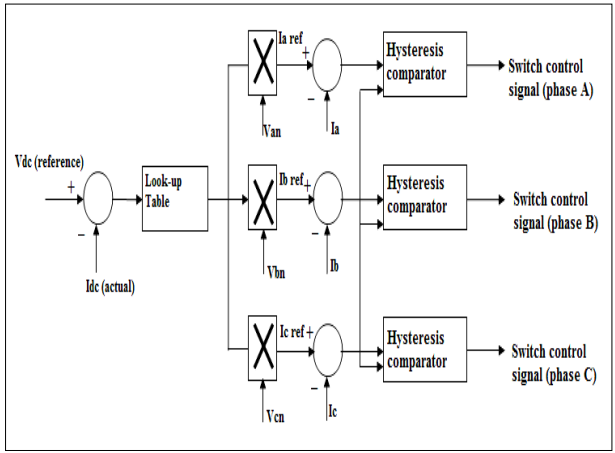

Fig.4. Block diagram of hysteresis current control VSC

The error of each phase current is controlled by a two level hysteresis comparator, which is shown in Fig.4. A switching logic is necessary because of the coupling of three phases. In order to explain the control method the mathematical equations should be introduced (Fig.5)

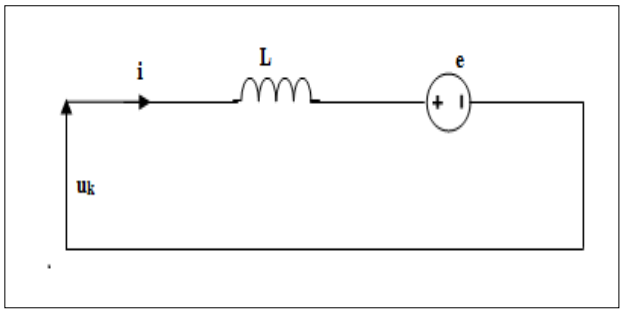

Fig.5. the load presentation

$$\frac{di}{dt} = \frac{1}{L}(u_k - e) \qquad (15)$$

According to equation (14), the current error deviation is given by,

$$\frac{di_e}{dt} = \frac{di}{dt} - \frac{di_{ref}}{dt} \qquad (16)$$

From equations (15) and (16) we have:

$$\frac{di_e}{dt} = \frac{1}{L}(u_k - u_{ref}) \qquad (17)$$

Where the reference voltage is the voltage which would allow that the actual current is identical with its reference value

### IV.3. DC/DC Converter Controller.

The maximum power tracker will generate a reference voltage that will be used to control the dc voltage at the rectifier dc side terminals. The dc-to-dc converter uses a simple feedback controller. The dc voltage reference is compared to the actual dc voltage, and the error signal is fed to a PI controller. The output signal is compared with a fixed frequency repetitive triangular waveform to deliver a signal that will turn ON or OFF the switch. This is shown in Fig.6. In many applications, a DC/DC Converter is used to produce a regulated voltage or current, derived from an unregulated power supply, or from a battery.

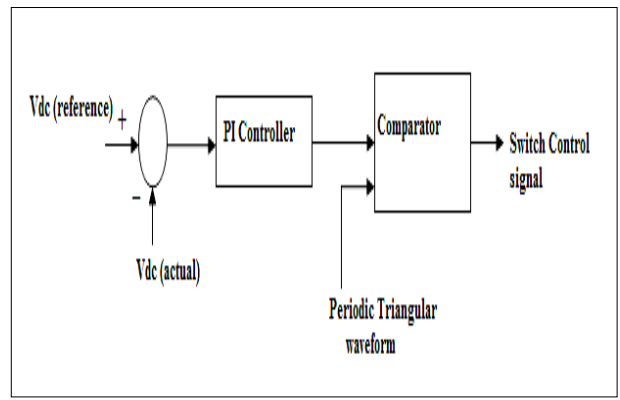

Fig.6. Block diagram of a typical DC-DC converter controller

## V. Simulation Results

The MATLAB-SIMULINK model of the WECS and the control systems are first presented. Simulation results with the MPPT in addition to the comments will then be given for two wind speeds. Simulation results of the WECS with MPPT algorithm are shown in Fig. 7 & Fig. 8. The simulation results consists of generator output for variable speeds 8 & 10 m/s, hysteresis current controlled VSC for variable speeds 8 to 10 m/s and DC-link voltages for variable speeds 8 to 10 m/s. At t = 10 s, wind speed is changed from 8 to 10 m/s in step, whereas tip-speed ratio is maintained at Cp maximum in steady state conditions. It is noticed that the controller is able to search for maximum power and keep the power coefficient of the wind turbine very close to its maximum.

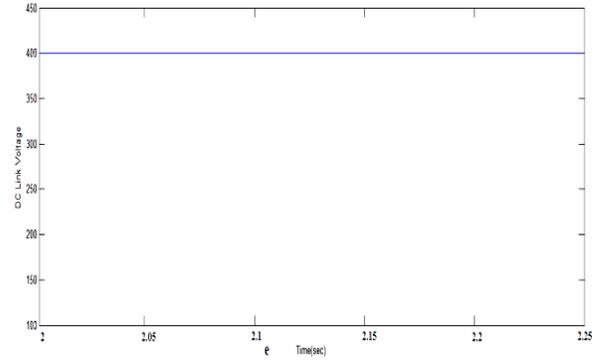

Fig.7. Simulation results for wind speed = 8 m/s

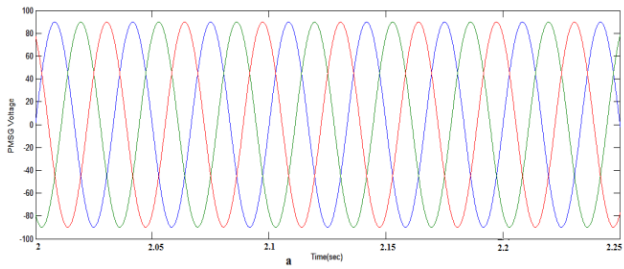
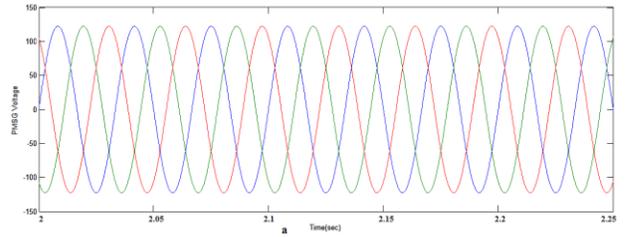
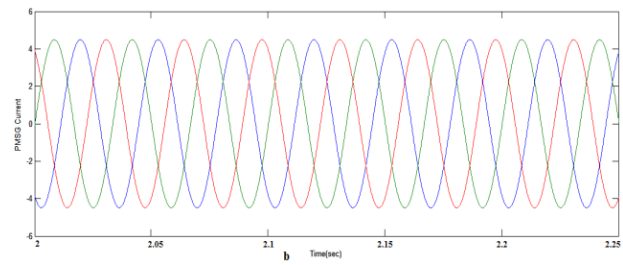
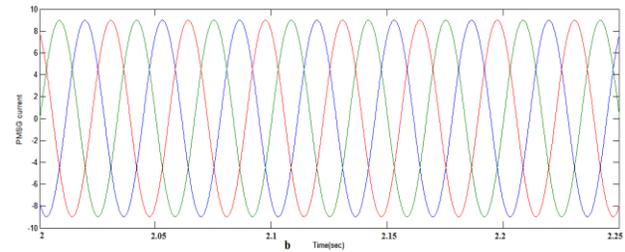
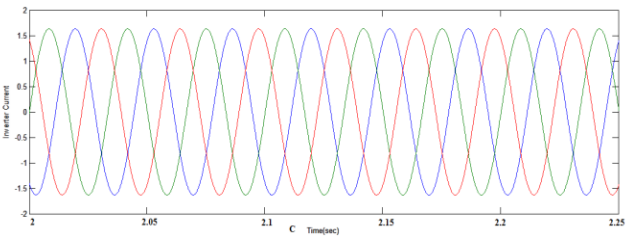
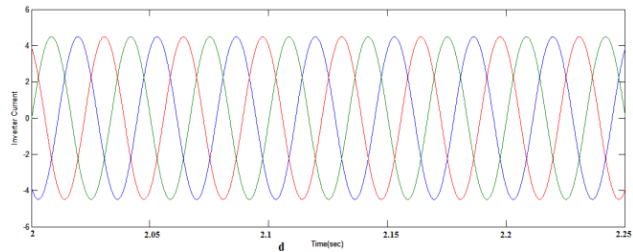
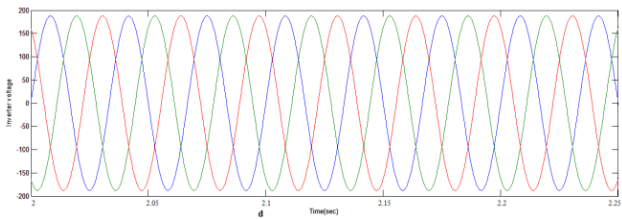
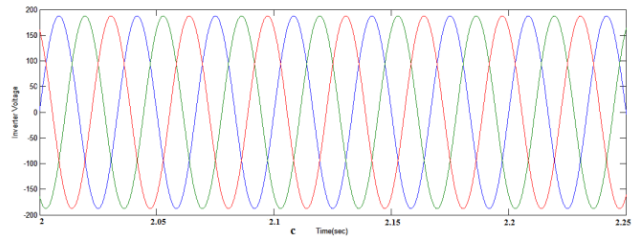

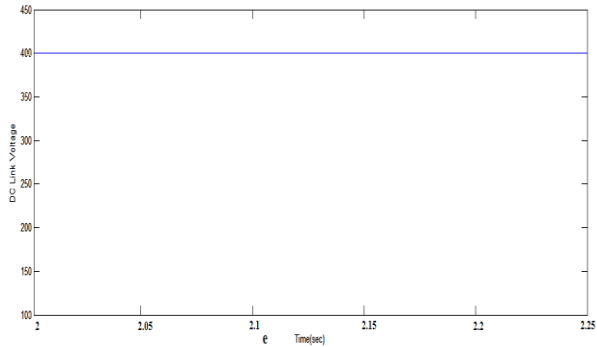

Fig.8. Simulation results for wind speed = 10 m/s

## VI. Conclusion

In WECS, The wind generator is formed by a fixed pitch wind turbine, a permanent magnet synchronous generator, a passive rectifier, a DC-DC boost converter and a hysteresis current controlled Voltage Source Converter (VSC). The maximum output power was extracted by a simple step and search control strategy. By converting DC output to AC, the isolated local load can be connected with hysteresis current controlled VSC. It is noticed that the controller is able to search for maximum power and keep the power coefficient of the wind turbine very close to its maximum.

## Authors' information

[1]Venkatachalam. Karthikeyan
Department of ECE, SVSCE, Coimbatore, India
karthick77keyan@gmail.com
[2] V.Jaganathan. Vijayalakshmi
Department of EEE, SKCET, Coimbatore, India
vijik810@gmail.com
[3] P.Jeyakumar
Department of ECE
Karpagam University, Coimbatore, India
Jeyak522@gmail.com

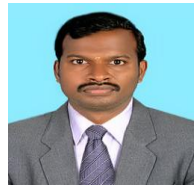

**Venkatachalam. Karthikeyan** has received his Bachelor's Degree in Electronics and Communication Engineering from PGP college of Engineering and technology in 2003 Namakkal, India. He received Masters Degree in Applied Electronics from KSR college of Technology, Erode in 2006. He is currently working as Assistant Professor in SVS College of Engineering and Technology, Coimbatore, He has about 7 years of teaching Experience.